\documentclass[aps,prb,twocolumn,superscriptaddress,10pt]{revtex4-1}
\setcitestyle{numbers,square}
\pdfoutput=1

\usepackage[utf8]{inputenc}
\usepackage[T1]{fontenc}
\usepackage[english]{babel} 
\usepackage{amsmath, amsfonts, amssymb}
\usepackage[locale=US, separate-uncertainty=true, range-units=single]{siunitx}
\usepackage{xspace}
\usepackage{nicefrac}
\usepackage{multirow}
\usepackage{physics}
\usepackage{graphicx}
\usepackage{enumerate}
\usepackage[dvipsnames]{xcolor}
\usepackage{hyperref}
\usepackage{float}
\hypersetup{
	colorlinks = true,
	linkcolor = blue,
	citecolor = blue,
	filecolor = blue,
	urlcolor = blue
}

\usepackage{prettyref}%
\newcommand{\pref}[1]{\prettyref{#1}}%
\newrefformat{fig}{Fig.~\ref{#1}}%
\newrefformat{tab}{Table~\ref{#1}}%
\newrefformat{sec}{Sec.~\ref{#1}}%
\newrefformat{app}{App.~\ref{#1}}%
\newrefformat{eq}{Eq.~(\ref{#1})}%

\newcommand{\SVO}{SrVO$_3$\xspace}

\begin{document}

\title{Training biases in machine learning for the analytic continuation of quantum many-body Green's functions}
\author{Rong Zhang}
\affiliation{Materials Theory, ETH Z\"u{}rich, Wolfgang-Pauli-Strasse 27, 8093 Z\"u{}rich, Switzerland}
\author{Maximilian E. Merkel}
\email{maximilian.merkel@mat.ethz.ch}
\affiliation{Materials Theory, ETH Z\"u{}rich, Wolfgang-Pauli-Strasse 27, 8093 Z\"u{}rich, Switzerland}
\author{Sophie Beck}
\affiliation{Center for Computational Quantum Physics, Flatiron Institute, 162 5th Avenue, New York, NY 10010, USA}
\author{Claude Ederer}
\email{claude.ederer@mat.ethz.ch}
\affiliation{Materials Theory, ETH Z\"u{}rich, Wolfgang-Pauli-Strasse 27, 8093 Z\"u{}rich, Switzerland}

\date{\today}

\begin{abstract}
We address the problem of analytic continuation of imaginary-frequency Green's functions, which is crucial in many-body physics, using machine learning based on a multi-level residual neural network. We specifically address potential biases that can be introduced due to the use of artificially created spectral functions that are employed to train the neural network. We also implement an uncertainty estimation of the predicted spectral function, based on Monte Carlo dropout, which allows to identify frequency regions where the prediction might not be accurate, and we study the effect of noise, in particular also for situations where the noise level during training is different from that in the actual data. Our analysis demonstrates that this method can indeed achieve a high quality of prediction, comparable or better than the widely used maximum entropy method, but that further improvement is currently limited by the lack of true data that can be used for training. We also benchmark our approach by applying it to the case of \SVO, where an accurate spectral function has been obtained from dynamical mean-field theory using a solver that works directly on the real frequency axis.
\end{abstract}

\maketitle

\section{Introduction}

A central quantity in quantum many-body physics is the single-particle Green's function, which describes the propagation of particles (or holes) in an interacting many-body system~\cite{bruus_many-body_2004}. For systems of interacting electrons without any explicit time dependence, the imaginary part of the frequency-dependent Green's function is related to the \emph{spectral function}, $A(\omega)$, which can be measured, e.g., by techniques such as (angle-resolved) photoemission spectroscopy \cite{damascelli_angle-resolved_2003}. However, numerical methods that work at finite temperature typically calculate the Green's function in imaginary time, $G(\tau)$, where $\tau \in [0, \beta)$ and $\beta = 1/(k_\mathrm{B} T)$ is the inverse temperature. In this case, an analytic continuation to real frequencies is required to obtain the physically relevant spectral function.

In this work, we specifically focus on the construction of spectral functions from imaginary time Green's functions obtained from Quantum Monte Carlo (QMC) solvers used in the context of dynamical mean-field theory (DMFT)~\cite{georges_dynamical_1996}. Within DMFT, the problem of interacting electrons within a solid is mapped on the solution of a fictitious Anderson impurity model describing individual atomic sites exchanging electrons with their environment. The environment is modeled as a non-interacting bath which is related to the atomic sites through a self-consistency condition. Some of the most widely used numerical methods to solve the corresponding impurity problem are based on QMC and determine the Green's function in imaginary time~\cite{gull_continuous-time_2011}.

The spectral function can be linked to the imaginary-time Green's function through the following integral equation:
\begin{align}
G(\tau) = -\int d\omega \frac{e^{-\tau\omega}}{1 + e^{-\beta\omega}}A(\omega) \quad , \label{eq:forward_transform}
\end{align}
which has to be inverted to extract the spectral function for a given $G(\tau)$.
However, the inversion of \pref{eq:forward_transform} is ill-conditioned, which means that the result is extremely sensitive to the statistical noise in $G(\tau)$.
A variety of different methods have been used to tackle this problem. The most widely used approach is the maximum entropy method (MaxEnt)~\cite{skilling_maximum_1989, gubernatis_quantum_1991}, but also other methods have been suggested, based on, e.g., Pad\'e approximants~\cite{vidberg_solving_1977}, averaged spectrum methods~\cite{ghanem_average_2020}, stochastic optimization~\cite{mishchenko_diagrammatic_2000}, or Carathéodory theory~\cite{fei_analytical_2021}. 

Recently, alternative approaches using different machine learning algorithms have been explored, see, e.g., Refs.~\onlinecite{arsenault_projected_2017, yoon_analytic_2018, fournier_artificial_2020, song_analytic_2020, xie_analytic_2022, huang_learned_2022}.
These methods have been shown to achieve good performance, often better or at least comparable to MaxEnt.
However, one problem of using supervised learning to invert \pref{eq:forward_transform} is the lack of true simulation data, i.e., spectral functions for real physical systems that are obtained directly on the real frequency axis without having to resort to analytic continuation in the first place.
Therefore, one typically resorts to artificially generated spectral functions that fulfil certain criteria on how spectral functions for real physical systems should look like. These hypothetical spectral functions are then used for the supervised learning, by applying the forward transformation, \pref{eq:forward_transform}, to obtain the corresponding $G(\tau)$, and then train the machine learning architecture to get $A(\omega)$ back from this $G(\tau)$. However, this means that the final performance can depend crucially on the way the initial spectral functions are generated.

Ideally, the performance of any approach for analytic continuation should be quantified on actual simulation results. This was indeed done in Ref.~\onlinecite{yoon_analytic_2018} for an insulating \SVO monolayer with a spectral function that exhibits two rather broad peaks.
However, since the corresponding Green's function was obtained in imaginary time using QMC, this only allows to compare the obtained result to other analytic continuation methods, e.g., MaxEnt, but not to quantify the performance relative to the ground truth, i.e., the correct real-frequency spectral function. 

Here, we use machine learning based on a multi-level residual neural network (NN) (described in \pref{sec:architecture}) to predict the spectral function corresponding to a given imaginary-time Green's function. Thereby, we put particular emphasis on cross-validating our NN on different data sets.
We also investigate the effect of different noise levels applied to the Green's function input data during training, in particular also the case where this noise level is different from that used for the testing.
Most importantly, we implement an uncertainty estimation, such that our approach does not only predict the most probable spectral function but also an uncertainty range for each frequency. This allows to highlight critical areas where the NN deems the prediction to be more difficult and also allows to estimate an empirical upper bound to the error relative to the ground truth.
Finally, we benchmark our approach on the spectral function obtained from a DMFT calculation of metallic bulk \SVO using a solver that works directly in real frequencies without having to resort to analytic continuation.

In the following, we first define how we generate our training data, how we represent our data and quantify the prediction error, and describe the NN architecture we use, before we discuss our results as outlined above. We also note that the corresponding code is available on GitHub \cite{rong_zhang_2022_6577294}. The code allows to reproduce the results of this paper as well as to predict general spectral functions obtained from DMFT calculations and is easily executable from GitHub through Google Colab.

\section{Methods}

As already outlined in the introductory section, we start by constructing large amounts of spectral functions, and then use the forward transformation, \pref{eq:forward_transform}, at a given inverse temperature $\beta$ to obtain the Green's function. We then add random noise to it, and train our NN with the noisy Green's function as input and the original spectral function as true output. Thereby, the deviation between predicted and true spectral function is used as cost function for training and accuracy metric for testing. We now discuss these steps in more detail.

\subsection{Generating and representing the data}
\subsubsection{Generation of initial data set}
\label{sec:data_recipes}

Similar to previous work~\cite{arsenault_projected_2017, yoon_analytic_2018, fournier_artificial_2020, xie_analytic_2022}, we use artificially generated spectral functions which are constructed from a superposition of Gaussian peaks:
\begin{align}
    A(\omega) = \sum_{i = 1}^{N_\mathrm{tot}} \left. \frac{c_i}{\sigma_i\sqrt{2\pi}} \exp\left(-\frac{(\omega-\mu_{i})^2}{2 \sigma_i^2} \right) \right/ \sum_{i = 1}^{N_\mathrm{tot}} c_i \quad . \label{eq:spec_func}
\end{align}
Here, $\mu_i$ and $\sigma_i$ are the center and spread of the $i$-th Gaussian peak with weight $c_i$, and the resulting spectral function is normalized to one.
This method allows to generate in principle infinitely many unique spectral functions which are then used 
for the training and testing of our NN. 

We limit the energy range of our spectral functions to $\omega \in [-\Omega_\mathrm{max}, \Omega_\mathrm{max}]$, where we choose $\Omega_\mathrm{max} = \SI{8}{eV}$ throughout this paper.
Then, inspired by Ref.~\onlinecite{arsenault_projected_2017}, we divide the energy window into different regimes: $[-\Omega_1, \Omega_1]$, $[-\Omega_2, \Omega_2]$, and $[-\Omega_3, \Omega_3]$ with \mbox{$0<\Omega_1<\Omega_2<\Omega_3<\Omega_\mathrm{max}$}.
In order to generate realistically looking spectral functions, we employ different constraints on the number, width, and weights of the Gaussian peaks in the different energy regions.

We do not generate any peaks in the outermost region, $\Omega_3<|\omega|<\Omega_\mathrm{max}$, to allow the spectral function to numerically vanish at $\pm \Omega_\mathrm{max}$. 
In the innermost region, $|\omega|<\Omega_1$, we restrict the number of peaks to a relatively small number and also generate peaks with higher average weights. This favors the formation of pronounced  quasi-particle-like features at low energies and simultaneously increases the probability to obtain some spectral functions with a gap around $\omega=0$.  
Furthermore, we use a larger average spread for peaks in the higher energy region, $\Omega_2<|\omega|<\Omega_3$, to avoid spectral functions with sharp features at such high energies.
As we will show later, these constraints allow us to generate a broad range of realistic looking spectral functions.

\begin{table}
    \centering
    \caption{Parameters corresponding to the two different ``recipes'' for generating spectral functions. Recipe 1 is used throughout most of the paper while recipe~2 is discussed only in \pref{sec:res_recipe2}.}
    \begin{ruledtabular}
    \begin{tabular}{lcc}
    & Recipe 1 & Recipe 2 \\ \hline
    \multicolumn{3}{l}{Frequency regimes (in eV)} \\[2pt] 
    $\Omega_1$, for number and weights & 1.6 & 0.96 \\
    $\Omega_2$, for spreads & 2.4 & 1.6 \\
    $\Omega_3$, for maximum of peak centers & 6.4 & 6.4 \\
    $\Omega_\mathrm{max}$, total frequency window & 8.0 & 8.0 \\\hline
    \multicolumn{3}{l}{Sampling range for number of peaks} \\[2pt]
    $N_\mathrm{tot}$, total & 8 - 34 & 4 - 6 \\
    For $|\omega| < \Omega_1$ & 0 - 4 & 0 - 3 \\ \hline
    \multicolumn{3}{l}{Sampling range for weights $c_i$} \\[2pt]
    For $|\omega| < \Omega_1$ & [0.5, 1] & [0.6, 1]\\
    For $|\omega| > \Omega_1$ & [0, 0.6] & [0, 0.2]\\ \hline
    \multicolumn{3}{l}{Sampling range for spreads $\sigma_i$ (in eV)} \\[2pt]
    For $|\omega| < \Omega_2$ & [0.12, 0.8] & [0.12, 0.8]\\
    For $|\omega| > \Omega_2$ & [0.48, 2.4] & [0.32, 0.8]\\
    \end{tabular}
    \end{ruledtabular}
    \label{tab:data_recipes}
\end{table}

The specific parameters used to generate spectral functions in this work are listed in Table~\ref{tab:data_recipes}. Generally, the number, position, width, and weight of the peaks in the different regions are randomly sampled from a homogeneous distribution in the specified ranges.
Furthermore, we require the spectral functions to vanish at $\pm\Omega_\mathrm{max}$, using a threshold value of \num{6.25e-5}.
If a spectral function does not vanish after uniform sampling of the parameters, the weight and spread of the Gaussian peak that contributes most to the spectral function at the boundary is reduced by a factor of 0.5 and 0.7, respectively. If this does not suffice, the whole peak is discarded. If required, this procedure is repeated until the spectral function vanishes at $\pm\Omega_\mathrm{max}$.

Note that we use two different parameter sets for generating spectral functions, as listed in \pref{tab:data_recipes}. ``Recipe~1'' is our main data set which is always used throughout this paper unless explicitly stated otherwise. It incorporates our best knowledge on the overall appearance of realistic looking spectral functions.
In contrast, ``Recipe~2'' is specifically designed to obtain spectral functions that also include sharper peaks at high absolute frequencies but with smaller spectral weights than for Recipe 1. This is motivated by our analysis of the spectral function of \SVO presented in \pref{sec:FTPS}. Recipe 2 is only used and discussed in \pref{sec:res_recipe2}.

\subsubsection{Extension of initial data set}

After generating an initial set of spectral functions in the way described above, we extend this data set to better account for linearity, scaling, and small variations in $A(\omega)$, as discussed in the following. 
This is an efficient technique to prevent overfitting and biases from the training data set. 

First, we note that \pref{eq:forward_transform} defines a linear relationship between $A(\omega)$ and $G(\tau)$.
To incorporate this property into the training set for the NN, we add two additional pairs of spectral functions for each originally generated spectral function. These pairs correspond to two different linear decompositions of $A(\omega)$ defined through \pref{eq:spec_func}, such that each additional spectral function contains a subset of peaks (with the correct overall normalization). 
We choose a decomposition into pairs of $N_\mathrm{tot}/2$ and $N_\mathrm{tot}/2$ as well as 1 and $N_\text{tot}-1$ number of peaks, which means that there are also single-peak spectral functions in the data set.

Second, according to \pref{eq:forward_transform}, the $G(\tau)$ corresponding to a scaled spectral function, $A(\omega) \rightarrow \alpha A(\alpha \omega)$ is mathematically identical to the Green's function corresponding to the original spectral function but transformed to the scaled temperature $\beta/\alpha$ (with appropriately scaled $\tau$ range).
While this does not correspond to the correct temperature dependence of the spectral function for a real physical system, incorporating this scaling relation of the kernel into our data set allows for a certain degree of temperature transferability, i.e., to obtain spectral functions at temperatures that are slightly different than the one used during the training of the NN. Furthermore, it adds spectral functions with different band widths.
We therefore add four additional spectral functions for each original $A(\omega)$, where the originally generated function is narrowed by a factor $\alpha$ of 1.2, 1.4, 1.6 and 1.8, respectively.

Finally, even small fluctuations in the spectral function often lead to variations in the Green's function that are larger than the typical QMC noise. This is particularly noticeable in the basis of Legendre coefficients introduced below for the first few Legendre coefficients.
Therefore, we add five more spectral functions for each $A(\omega)$ obtained after initial generation and extension for linearity and scaling. These additional spectral functions contain 10 more random Gaussian peaks with very small weights, $c_i \in [0.001, 0.01]$, such that the overall appearance of $A(\omega)$ is almost unaltered.

\subsubsection{Representation of data, noise, and error quantification} \label{sec:methods_representation}

For a given spectral function, we obtain the imaginary-time Green's function $G(\tau)$ at an inverse temperature $\beta =\SI{40}{eV^{-1}}$ from Eq.~\eqref{eq:forward_transform} and represent it in terms of Legendre polynomials~\cite{boehnke_orthogonal_2011}, leading to the basis of Legendre coefficients $G_l$, with typically \numrange{20}{50} coefficients. We use 80 coefficients as input to our NN, enough to safely include all information even for well-sampled QMC calculations, and expect the NN to learn which coefficients are relevant and which contain only noise.
The Legendre parametrization is notably more compact than a discretization in imaginary time or using Matsubara frequencies. 

Similar to previous works, we model the statistical QMC error as uniform, i.e., $l$-independent noise, sampled from a Gaussian distribution with variance $\sigma^2$, added to each $G_l$ coefficient.
We do not consider any correlations of the error between different Legendre coefficients, i.e., the corresponding covariance matrix $K_{ll'} = \operatorname E[(G_l - \operatorname E[G_l]) (G_{l'} - \operatorname E[G_{l'}])]$, where $\operatorname E$ is the expectation value, is an identity matrix scaled by $\sigma^2$. In \pref{sec:effect_noise}, we analyze the covariance matrix of a real DMFT-QMC calculation in order to validate these assumptions.
Furthermore, we initially use different noise levels for the training, $\sigma_\mathrm{train}$, and the testing, $\sigma_\mathrm{test}$, to address the robustness of the predicted spectral functions with respect to uncertainties in the noise estimation of the underlying data.

All spectral functions are represented on an equidistant frequency grid, $\omega_i$, with $N_\omega = 800$ points. This makes it straightforward to enforce positivity and normalization of the predicted spectral functions. For our purpose, these advantages outweigh the higher dimension of the basis compared to more compact representations such as conformal coefficients \cite{arsenault_projected_2017} or a basis obtained from singular value decomposition.

We quantify the quality of our prediction in terms of the mean absolute error (MAE), $\bar \varepsilon$, defined as the discretized integral of the pointwise absolute error, \mbox{$\varepsilon(\omega) = |A_\mathrm{true}(\omega) - A_\mathrm{predicted}(\omega)|$}, over the whole frequency range:
\begin{align}
    \bar \varepsilon = \int_{-\infty}^\infty \mathrm{d}\omega\, \varepsilon(\omega) \approx \frac{2\Omega_\mathrm{max}}{N_\omega - 1} \sum_{i=1}^{N_\omega} \varepsilon(\omega_i)  \quad . 
    \label{eq:mae}
\end{align}
Here, $A_\mathrm{true}(\omega)$ is the correct spectral function and $A_\mathrm{predicted}(\omega)$ the function predicted by the NN.
Our definition of the MAE has the advantage of being unitless and fulfills $\bar\varepsilon \in [0, 2]$.

\subsection{Neural network architecture and uncertainty estimation}\label{sec:architecture}

\begin{figure*}
    \centering
    \includegraphics[width=1\linewidth]{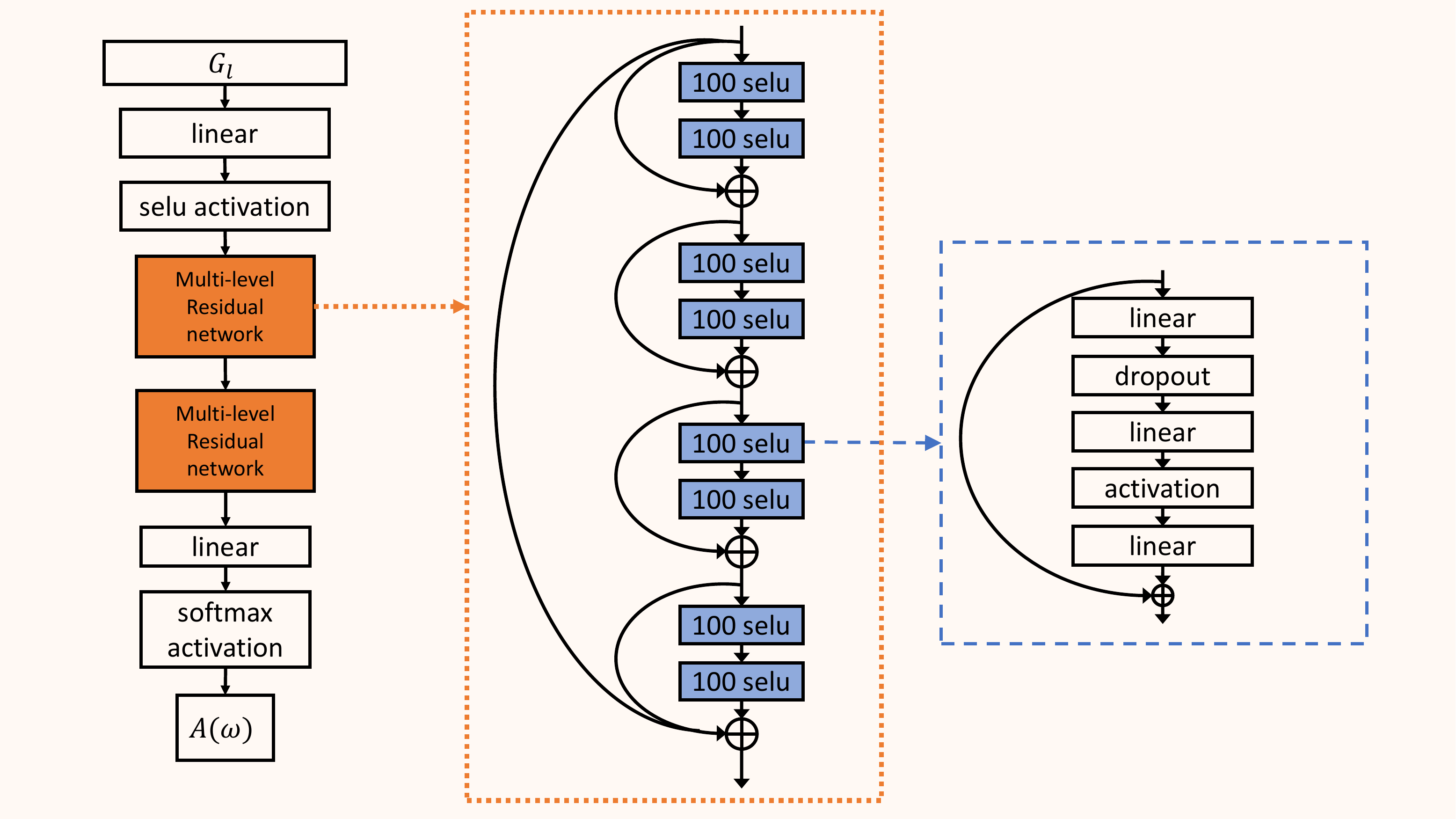}
    \caption{The NN architecture. The complete structure on the left, a more detailed view of the "multi-level residual block" in the center and the basic residual unit, labeled ``100 selu'', on the right. This network architecture allows for accurate training while suppressing saturation effects with a manageable number of trainable parameters.}
    \label{fig:resnet2222}
\end{figure*}

We use the python libraries TensorFlow and Keras to implement our NN~\cite{tensorflow2015-whitepaper, chollet2015keras}.
Initially, we tested different NN architectures
that have been used in previous work for the analytical continuation problem, such as a multilayer perceptron (MLP) model~\cite{fournier_artificial_2020}, a residual NN where a MLP model is used to model the difference between the target and input (\textit{cf.} Ref.~\onlinecite{he_deep_2016}), and the Adams-Bashforth ResNet~\cite{xie_analytic_2022}.
In the end, we decided to use the multi-level residual NN as shown in \pref{fig:resnet2222}, which resulted in the lowest training and validation error of all the architectures we have tried.

The right-most part of \pref{fig:resnet2222} shows the basic building block, i.e., the lowest level of residual learning of our multi-level residual NN. Each linear layer applies a linear transformation to the input vector of length 100. The dropout layer randomly sets \SI{10}{\%} of the elements of the input vector to zero in every step, which reduces overfitting and gives an estimation for the prediction uncertainty, as we describe below in more detail.
In the activation layer, the non-linear scaled exponential linear unit (selu) function, as implemented in Keras, is applied to the input. Finally, the original input is added to the block output, symbolized by the arrow bypassing the different layers.
The middle part of \pref{fig:resnet2222} shows how the residual blocks are concatenated. The additional connections in between make the NN residual on higher levels.

The left part of \pref{fig:resnet2222} shows the pre- and post-processing steps applied to the central block. The input $G_l$ of dimension 80 is first passed through a linear layer, which scale up the data to the dimension of 100 that the residual blocks use, and then through a selu activation function.
The output of this is then used in the two ``multi-level residual network'' blocks, which have been defined in the central panel. A final linear layer then scales the dimension up to the number of points on the $\omega$ grid, and in the last step, the softmax activation function ensures proper normalization of the resulting spectral function.

The higher-level residual connections reduce the saturation effects observed in the Adams-Bashforth ResNet when increasing the number of residual blocks.
Moreover, our architecture can achieve a greater depth with a comparable or even smaller number of parameters to be tuned in the learning process, containing only 100 perceptrons per layer in the residual blocks.

The stochastic gradient descent is performed with the ADAM optimizer \cite{kingma_adam_2017}. Starting from a learning rate of \num{5e-5}, the learning rate is reduced by a factor of 1.2 whenever the validation loss stops decreasing, until a threshold learning rate of \num{5e-7} is reached and the training is stopped.

\textit{Monte-Carlo (MC) dropout} is an efficient way to quantify the prediction uncertainty \cite{gal_dropout_2016}. In our NN, this is implemented through the dropout layer in the ``100 selu'' block. Therefore, by performing 100 prediction runs with our NN on the same input $G_l$, we obtain both an average spectral function $A(\omega)$ as well as a standard deviation $\sigma_\mathrm{dropout} (\omega)$ for each frequency $\omega$.
However, the standard deviation $\sigma_\mathrm{dropout}(\omega)$ obtained in this way is known to underestimate the absolute error $\varepsilon (\omega)$ between predicted and true output. Ref. \onlinecite{gal_dropout_2016} argues that the uncertainty of dropout models depends on the activation function and NN topology, which makes calibration necessary. 
In \pref{sec:res_uncertainty}, we therefore carefully analyze the quality of the uncertainty estimation based on $\sigma_\mathrm{dropout}$.

\subsection{Other computational details}
\label{sec:other_details}

We also benchmark the predictions from our NN against the state-of-the-art maximum-entropy package MaxEnt, contained in the TRIQS library (version 1.0.0) \cite{parcollet_triqs_2015, maxent}.
Thereby, we use the $\chi^2$-curvature analyzer to find the optimal regularization parameter $\alpha$ with a constant error parameter for $G(\tau)$ of 0.03
and a hyperbolic mesh in the range of \SIrange{-16}{16}{eV} with 800 points.
For a direct comparison with the NN on consistent test data, we first transform $A(\omega)$ into $G_l$, then add the noise, and finally transform the result into imaginary time to generate the input required for MaxEnt.

To validate our model for the noise on $G_l$, we use density-functional theory (DFT) plus DMFT calculations for \SVO to obtain a realistic covariance matrix $K_{ll'}$ corresponding to the local Green's function obtained from a QMC solver.
For that, we first converge the DFT+DMFT calculation. Starting from this converged solution but using a different random seed, 200 additional independent QMC runs are performed. With the resulting local Green's functions $G_l$ of the V cation, we calculate $K_{ll'}$ as defined in \pref{sec:methods_representation}.

Furthermore, we also calculate the Green's function of SrVO$_3$ within DFT+DMFT directly on the real frequency axis, by using the recently developed solver based on fork tensor-product states (FTPS)~\cite{bauernfeind_fork_2017}. This serves as a final benchmark of our NN by comparing the spectral function obtained without analytic continuation with the spectral function predicted from the corresponding $G_l$, which are obtained from the forward-transform, \pref{eq:forward_transform}, and subsequent conversion to Legendre coefficients.
\SVO is a standard example for a correlated metal with a relatively simple crystal structure that is straightforward to treat within DFT+DMFT, due to its well-separated subset of partially filled V-$t_{2g}$ bands.
Further details of these DFT+DMFT calculations are described in \pref{app:dft_dmft}.

\section{Results}

\subsection{Influence of the noise level}\label{sec:effect_noise}

\begin{figure}
    \centering
    \includegraphics[width=1\linewidth]{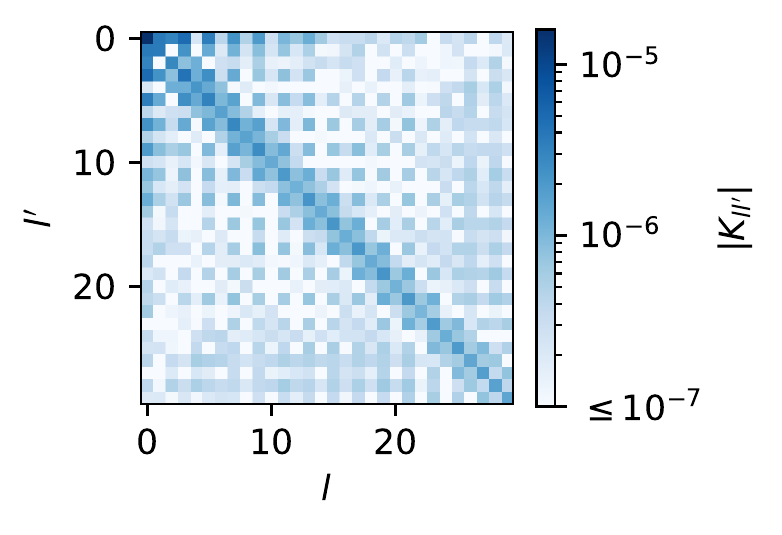}
    \caption{The covariance matrix $K_{ll'}$ corresponding to the imaginary-time Green's function in the Legendre basis, $G_l$, for $l, l' < 30$, evaluated from 200 independent QMC runs on top of a pre-converged DFT+DMFT calculations for \SVO.}
    \label{fig:res_noise_distr_qmc}
\end{figure}

Before we analyze the effect of the noise level applied to $G_l $ on the quality of prediction of our NN, we first test the validity of our assumption of an $l$-independent Gaussian-distributed noise with no correlations between different $l$.
As described in \pref{sec:other_details}, we obtain a realistic covariance matrix $K_{ll'}$ by sampling over 200 independent QMC runs on top of a pre-converged DFT+DMFT calculation for \SVO. The result is visualized in \pref{fig:res_noise_distr_qmc}.

One can see that the covariance matrix is largest along the diagonal, while the off-diagonal elements decay relatively quickly away from the diagonal. The diagonal terms are indeed almost constant with $K_{ll} \approx \num{2e-6}$, except for $l=0$ with $K_{00} \approx \num{2e-5}$. However, particularly for small $l, l'$ the off-diagonal elements are not negligible. 
As discussed in \pref{sec:methods_representation}, our model of uniform, independent noise on the Legendre Green's function corresponds to a covariance matrix $K_{ll'} = \delta_{ll'} \sigma^2$. 
Thus, our analysis shows that, while this model of the noise is not perfect, it nevertheless captures the most important features of the covariance matrix.
Furthermore, except for $l=0$, this model fits best to the calculated covariance matrix for a noise level of $\sigma \approx 0.0014$.
In the following, we will therefore consider a noise level of $\sigma = 10^{-3}$ as a typical value for well-converged QMC data.
We note that constructing a more complex model for the noise is far from trivial, and that the more detailed features of the covariance matrix can also depend on the specific sampling method used in the QMC calculations. Therefore, a more complex model might also become less general.

\begin{figure}
    \centering
    \includegraphics[width=1\linewidth]{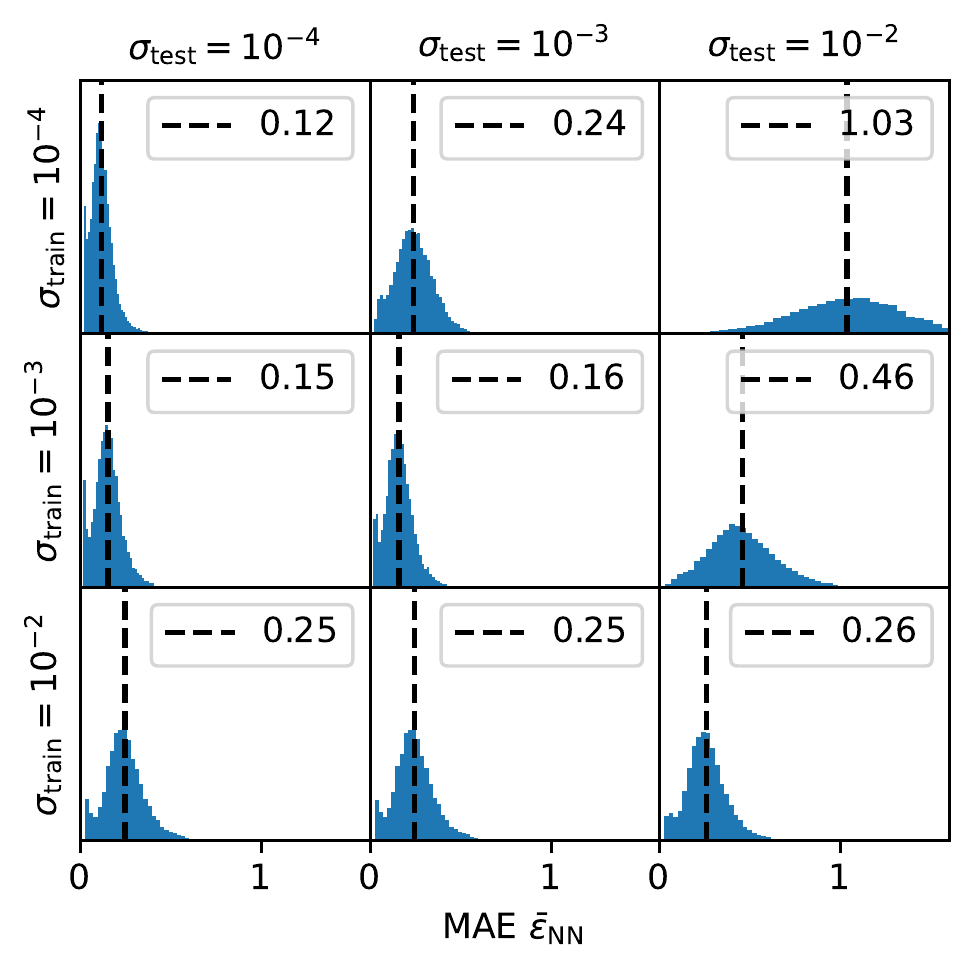}
    \caption{Distribution of the MAE corresponding to the predicted spectral functions with different noise levels applied to the training and testing data, $\sigma_\text{train}$ and $\sigma_\text{test}$, respectively. The histogram in blue represents the distribution of the MAE, while the dashed line shows its average over all test spectral functions. The value of the average MAE is also indicated in the legends of the individual panels.}
    \label{fig:res_effect_noise}
\end{figure}

Next, we investigate how the noise level in the data affects the quality of prediction of our NN.
Thereby, we also consider the case where the noise level in the training data ($\sigma_\text{train}$) is different from the noise level in the data used for testing ($\sigma_\text{test}$).
This is particularly crucial if the exact noise level of the data is unknown or hard to determine accurately, which is often the case for data obtained from real simulations.
The results are summarized in \pref{fig:res_effect_noise}, which shows histograms of the MAE of the predicted spectral functions for the various cases with different $\sigma_\text{train}$ and $\sigma_\text{test}$.

It can be seen that the minimal MAE for a specific noise level in the test data is always obtained for the NN that was trained with identical noise level, and that the corresponding MAE increases with increasing noise level. Furthermore, the quality of prediction deteriorates significantly if the NN was trained at a low noise level but is then used on data with a much higher noise level (e.g., $\sigma_\text{train}=10^{-4}$ and $\sigma_\text{test}=10^{-2}$). On the other hand, a NN that was trained with a higher noise level, leads to a very consistent MAE for all $\sigma_\text{test} \leq \sigma_\text{train}$.

This can easily explained as follows. If the testing noise level is higher than during the training, the NN interprets the noise in the testing data as features and thus arrives at inaccurate results. In turn, if the testing noise level is smaller than during the training, the NN takes information as noise, therefore failing to resolve some features that it would be able to resolve with lower training noise, but still gives the same quality of prediction as for the original higher noise level.
Therefore, it is clear that, if in the real application the noise level is not known exactly, it is better to use a relatively generous estimate. 

\subsection{Comparison to MaxEnt}
\label{sec:res_maxent}

\begin{figure}
    \centering
    \includegraphics[width=1\linewidth]{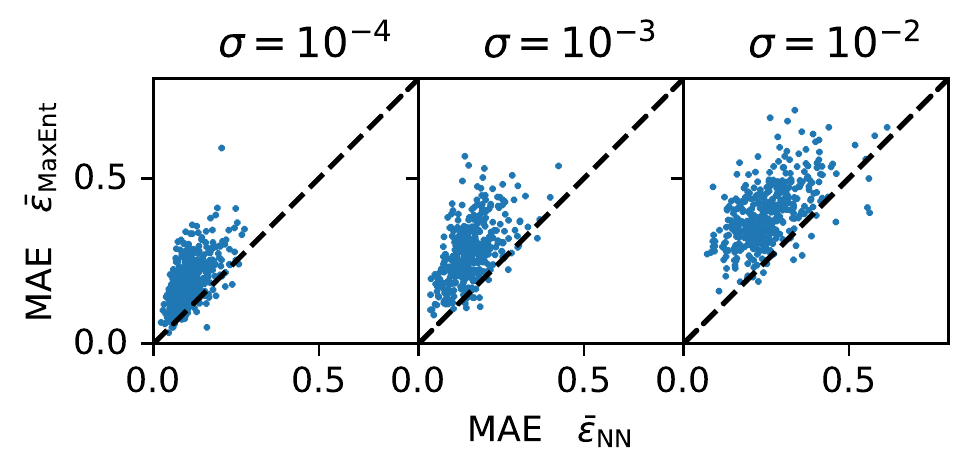}
    \caption{Comparison of the MAE for 400 spectral functions predicted by the NN ($\bar{\varepsilon}_\text{NN}$) and by MaxEnt ($\bar{\varepsilon}_\text{MaxEnt}$), respectively, for different noise levels, $\sigma$, applied to both testing (for the NN) and training data.}
    \label{fig:res_comp_maxent}
\end{figure}

Next, we compare the performance of our NN against the well established and widely used MaxEnt approach.
Thereby, we always use the NN trained with a noise level identical to that in the testing data, and for each noise level we use the same set of 400 test spectral functions (generated from the original data recipe without data set extension) for both the NN and MaxEnt.

In \pref{fig:res_comp_maxent}, we compare the MAE obtained with MaxEnt with that obtained from the NN prediction. For all considered noise levels, the large majority of all points in these scatter plots lie above the diagonal, indicating a larger error of prediction of MaxEnt compared to the NN.
Furthermore, the few data points where the NN error is larger than that of MaxEnt are all relatively close to the diagonal, which means that there are only very few spectral function where MaxEnt drastically outperforms our NN.
We therefore conclude that our NN performs much better on this data set than MaxEnt, which is a very promising result, in line with the literature \cite{arsenault_projected_2017, fournier_artificial_2020}.

However, one should also note that this comparison is somewhat biased, since the NN has been specifically optimized for spectral functions obtained with this data recipe during the training, whereas MaxEnt works for general cases. Depending on the generality of the data recipe, this can either be viewed as a negative feature, indicating overfitting of the NN, or as a positive feature, showing that the NN successfully learns about the general shape of spectral functions.
We will again address this question from \pref{sec:res_unknown_data} onward, but will first discuss the uncertainty quantification based on MC dropout in the following section.

\subsection{Uncertainty estimation} \label{sec:res_uncertainty}

From now on, except where otherwise noted, we always use a noise level of $\sigma_\mathrm{train} = \sigma_\mathrm{test} = 10^{-3}$, unless explicitly stated otherwise. Note that this is indeed close to our estimation of a realistic magnitude of the QMC noise in DMFT calculations, based on the covariance matrix analyzed in \pref{fig:res_noise_distr_qmc}. 

\begin{figure}
    \centering
    \includegraphics[width=1\linewidth]{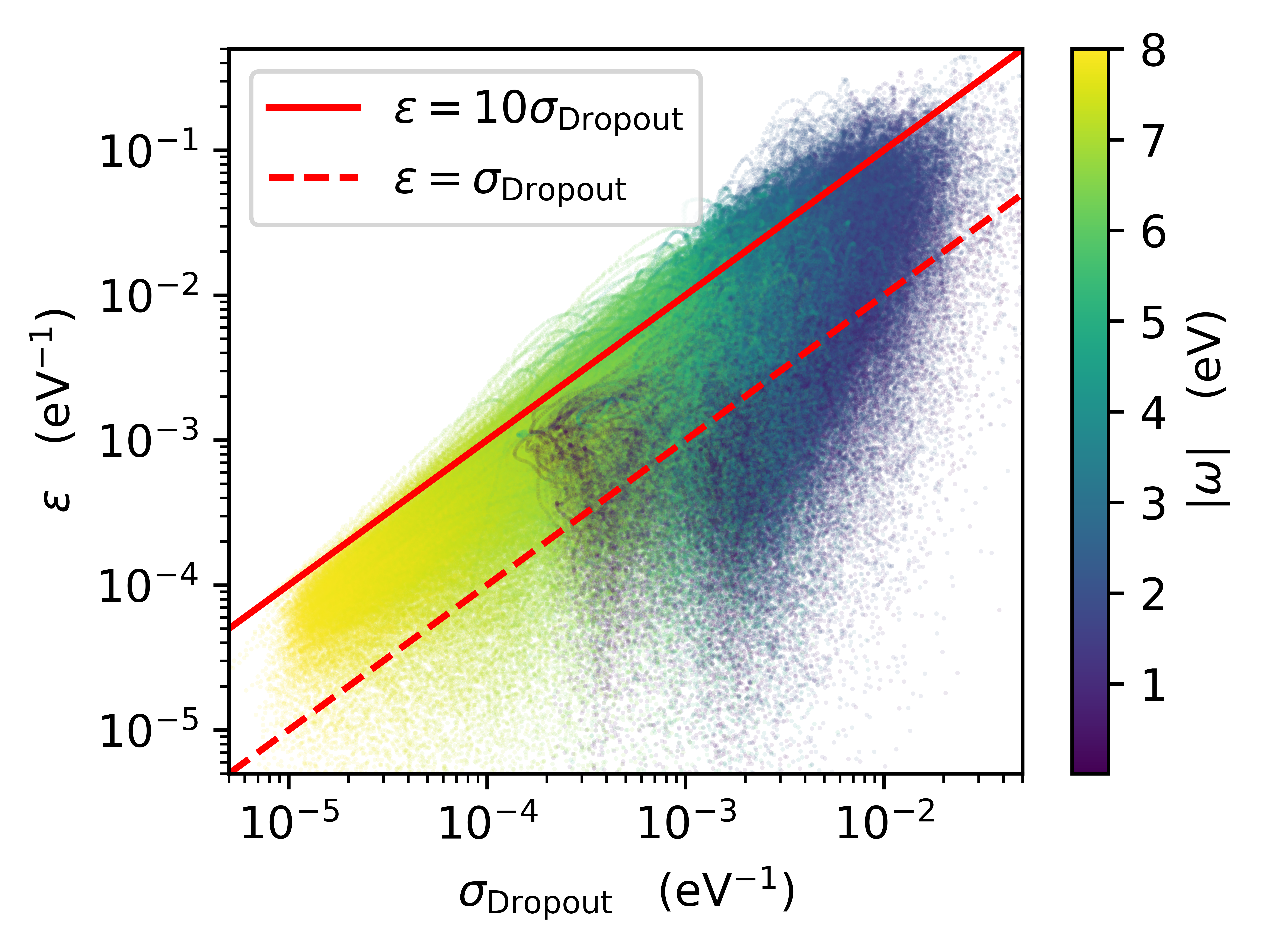}
    \caption{The point-wise absolute error $\varepsilon(\omega)$ plotted over the standard deviation from the MC dropout, $\sigma_\text{Dropout}$ for a noise level of $\sigma=10^{-3}$.
    The dashed line represents the case where $\varepsilon = \sigma_\text{Dropout}$, whereas the solid line, $\varepsilon=10\,\sigma_\text{Dropout}$, represents the value used for our empirical uncertainty estimation.}
    \label{fig:res_uncert_quant}
\end{figure}

\pref{fig:res_uncert_quant} shows the relation between the MC-dropout standard deviation $\sigma_\mathrm{Dropout} (\omega)$ and the absolute pointwise error $\varepsilon(\omega)$, evaluated over a set of 1000 spectral functions generated independent from the training set.
The color of each data-point indicates the absolute value of the corresponding frequency $\omega$.
One recognizes that larger errors tend to occur for low to medium absolute frequencies $|\omega| < \SI{5}{eV}$, whereas high absolute frequencies are typically well described. The latter is due to the fact that all the spectral functions we generate are forced to decay towards $|\omega|=\Omega_\mathrm{max}$.

Furthermore, one can see that a small dropout standard deviation seems to always correspond to a small error, whereas a large standard deviation can be associated with anything from a small to a very large absolute error.
Thus, even though there seems to be some correlation between $\varepsilon$ and $\sigma_\text{Dropout}$, a simple relation such as $\varepsilon(\omega) = \sigma_\mathrm{Dropout} (\omega)$ (dashed red line) does not hold. Nevertheless, it appears that $\sigma_\text{Dropout}$ allows to estimate an empirical approximate upper bound for the absolute pointwise error according to $\varepsilon(\omega) \leq 10 \, \sigma_\mathrm{Dropout} (\omega)$, as indicated by the solid red line in Fig.~\ref{fig:res_uncert_quant}. This estimation holds true for \SI{92}{\%} of the data points in the plot, where most of its violations occur at medium frequencies $|\omega| \approx \SI{4}{eV}$.
Comparing different noise levels, it appears that the reliability of this uncertainty estimation tends to increase for smaller noise levels, see \pref{tab:crossval_uncertainty} and \pref{app:uncertainty}.

In the following we will empirically use $10\,\sigma_\text{Dropout}$ to indicate the uncertainty of the NN prediction. We note that there is no mathematical guarantee that this works equally well for unknown Green's functions. However, if the input Green's function is similar to our testing data, it is reasonable to expect that the uncertainty estimation is qualitatively correct. Additionally, a large uncertainty serves as a good indicator to identify potentially problematic features in a certain frequency range. 
This will be further analyzed in the remaining sections.

\subsection{Examples of predicted spectral functions}

\begin{figure}
    \centering
    \includegraphics[width=1\linewidth]{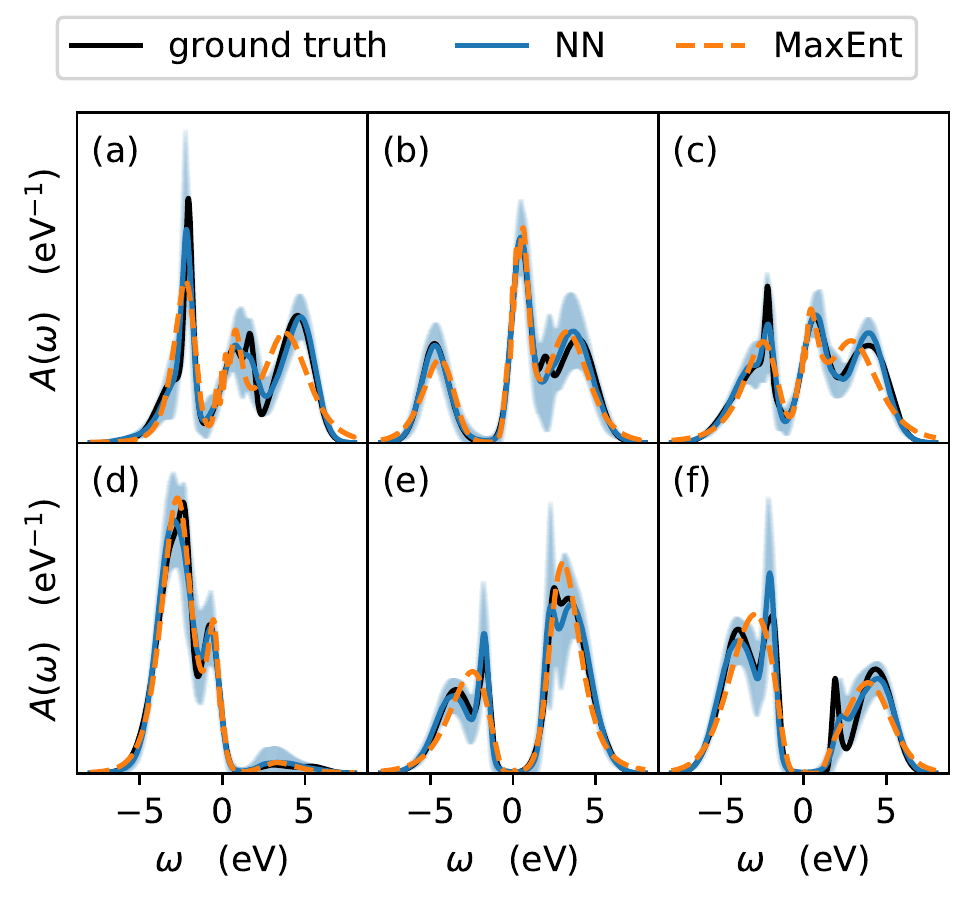}
    \caption{Examples of spectral functions constructed from data recipe 1. Each panel contains the ground truth, the NN prediction with the estimated uncertainty range, and the MaxEnt prediction. Predicted spectral functions are obtained with a noise level of $\sigma=10^{-3}$ on $G_l$.}
    \label{fig:res_examples_spectral}
\end{figure}

\pref{fig:res_examples_spectral} shows examples of spectral functions randomly chosen from our data set, together with the predictions of our NN (including uncertainty estimation) and MaxEnt, based on a noise level of $10^{-3}$. 
Three of the spectral functions shown in \pref{fig:res_examples_spectral} correspond to ``metallic'' cases with approximate half-filled bands [(a)-(c)], one represents an almost completely filled case [(d)], and two of them are gapped around $\omega=0$ [(e) and (f)], which would indicate an insulating system. Thus, \pref{eq:spec_func} with parameters chosen according to ``recipe 1'' from \pref{tab:data_recipes} indeed leads to large variety of diverse and non-trivial spectral functions.

Both MaxEnt and the NN are able to correctly describe the main features of all spectra, i.e., metallic versus insulating character, the main peak structure, and band-width. However, MaxEnt clearly fails to resolve more detailed features such as the double peak structures both above and below the gap in (e) and (f), or the shoulders towards low energies in (a) and (c).
These deficiencies appear mostly if two smaller peaks are rather close to each other or far away from zero energy. The merging of peaks can then also lead to a shift of the peak centers, a misestimation of peak heights and width, and problems with the decay towards high absolute frequencies [see, e.g., (a) and (c)].

Overall, the NN appears clearly superior to MaxEnt, although it also fails to resolve very detailed features such as the two peak structure close to zero energy in (a), or the small and narrow peak around 2\,eV in (b). It also slightly underestimates the heights of the very sharp peaks in (c) and (f), but is able to resolve all four peaks in (f) as well as the low energy shoulder in (c). 
In all cases, the uncertainty estimation works well in identifying critical parts of the prediction and even captures almost all features within the predicted range.
In conclusion, the NN is more reliable for these examples than MaxEnt, which agrees with the above comparison of the MAE between the NN and MaxEnt.

\subsection{Applying the NN to untypical data} \label{sec:res_unknown_data}

\begin{figure}
    \centering
    \includegraphics[width=1\linewidth]{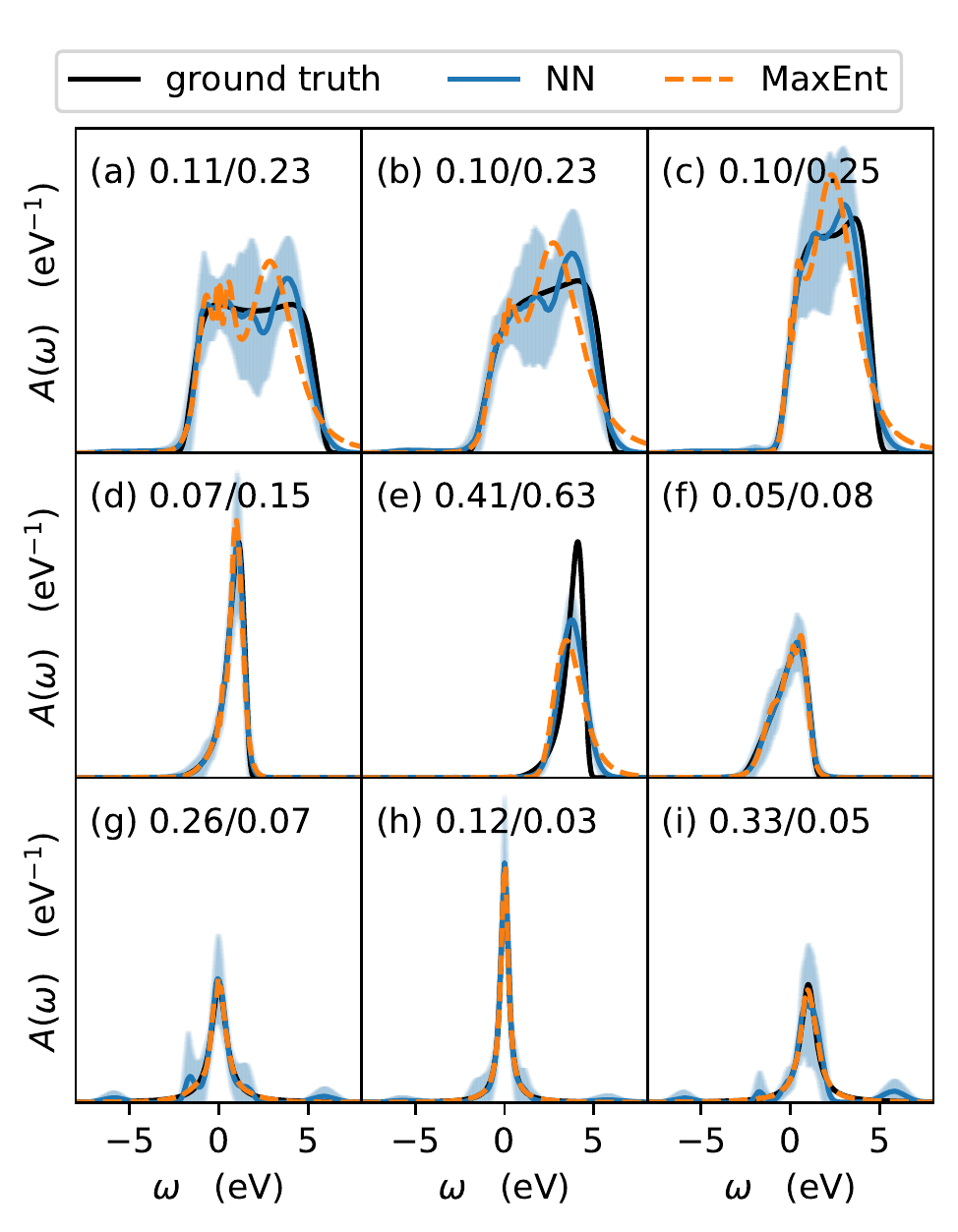}
    \caption{Comparison between the NN and MaxEnt predictions on a special set of spectral functions not represented by \pref{eq:spec_func}, for a noise level of $\sigma=10^{-3}$. Cases (a)-(c) correspond to wider plateaus, cases (d)-(f) to asymmetric peaks, and (g)-(i) to Lorentzian peaks. The values in the top of each figure indicate the corresponding MAE for NN and MaxEnt, respectively.}
    \label{fig:res_special_spectral}
\end{figure}

In the previous sub-sections, in particular in the comparison wit MaxEnt, we have come across the question of how versatile our recipe for data generation and thus the resulting trained NN really is. To further investigate a potential bias of our approach, we now apply both the NN and MaxEnt to spectral functions that we specifically designed to be different from those included in the training data for the NN. 
\pref{fig:res_special_spectral} shows a selection of such spectral functions, which contain plateaus, asymmetric peaks, and Lorentzians with a different asymptotic decay as the Gaussians used in \pref{eq:spec_func}. Such features cannot easily be expressed by \pref{eq:spec_func}, and thus represent a hard challenge to our NN. 

Starting with the plateaus in the top row, \pref{fig:res_special_spectral}(a)-(c), 
one sees that the flat parts of the plateaus lead to oscillations in the MaxEnt prediction and also, to a lesser extent, in that of our NN. For the NN, this clearly reflects the bias towards the data seen during the training, i.e., the NN tries to fit these parts with a superposition of Gaussian peaks. Additionally, the uncertainty of our NN is very large in the plateau area, and mostly includes the ground truth. This behavior indicates that plateaus are fundamentally hard to resolve. 

The middle row, panels (d)-(f) in \pref{fig:res_special_spectral}, contains asymmetric peaks, centered either at $\omega=0$ or off-centered, and with different widths.
Both MaxEnt and the NN perform rather well for the centered peaks, while the off-center peak in (e) appears to be more difficult, with the NN doing a slightly better job than MaxEnt. 
The uncertainty quantification works well for the centered peaks but vastly underestimates the deviations for the off-center peak, not clearly identifying this as a critical case.

The bottom row, panels (g)-(i) in \pref{fig:res_special_spectral}, contains Lorentzian peaks.
Here, we see that MaxEnt does not have a problem with these peaks, representing them almost perfectly, while the NN tries to fit the spectral functions with Gaussians, leading to spurious high-energy peaks in panels (g) and (i). The Lorentzian peaks are also the only spectral functions in \pref{fig:res_special_spectral} where MaxEnt outperforms the NN in terms of the MAE.
However, in all three cases, the uncertainty quantification of the NN correctly identifies the sides of the peak as well as the high energy features as problematic areas.

In summary, the NN seems to outperform MaxEnt even for cases that are very different from the training data, such as wide plateaus and asymmetric peaks, indicating the versatility of this approach. It also appears that in some cases the NN exhibits a certain bias towards predicting something that is more similar to the training data.  
However, except for case (e), the uncertainty estimation works rather well in identifying the regions where the prediction might not be accurate.

\subsection{Testing on \SVO from real-frequency DMFT}
\label{sec:FTPS}

\begin{figure}
    \centering
    \includegraphics[width=1\linewidth]{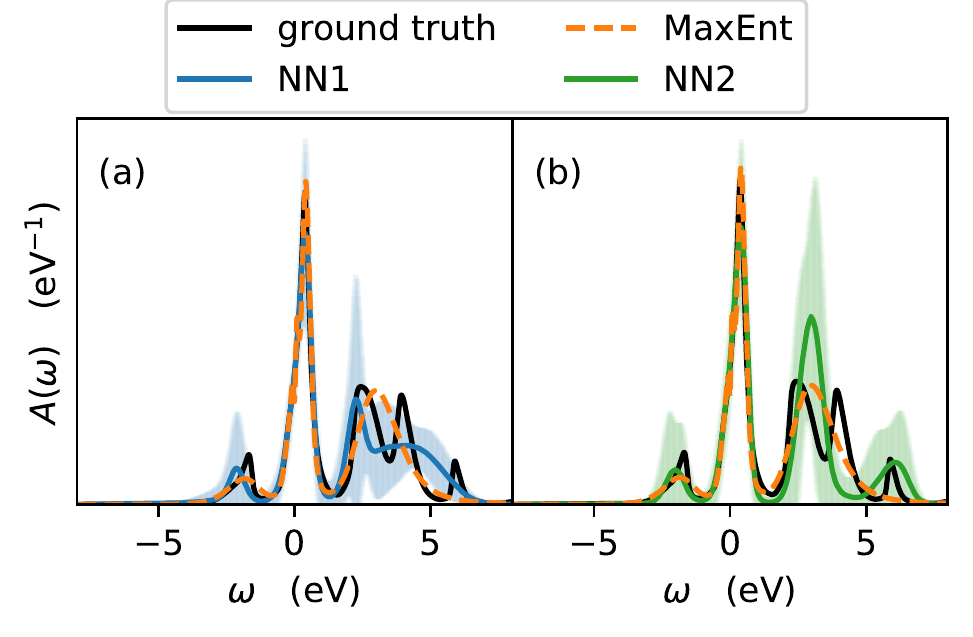}
    \caption{Spectral functions obtained with the NN trained on (a) data recipe 1 (NN1) and (b) data recipe 2 (NN2), together with the MaxEnt prediction for the Green's function obtained from a DFT+DMFT calculation for \SVO using the FTPS solver (with an artificial noise of $10^{-3}$ added to the forward-transformed FTPS data). The ground truth is the result obtained by the FTPS solver without analytic continuation. 
    }
    \label{fig:res_ftps_spectral}
\end{figure}

Next, we benchmark our method on real simulation data, using the spectral function obtained for \SVO from a DMFT calculation with the FTPS solver, an impurity solver based on tensor networks which works directly on the real-frequency axis and therefore does not require analytic continuation. Since we use the FTPS spectral function as another test function for the NN, we apply the same procedure to the $A(\omega)$ obtained from the FTPS solver as for the artificial spectral functions obtained from \pref{eq:spec_func}. Thus, we construct $G(\tau)$ from the forward transformation, \pref{eq:forward_transform}, add noise to the corresponding $G_l$ coefficients, let the NN predict the spectral function from that, and then compare with the original ground truth.
We again use $\beta = \SI{40}{eV^{-1}}$ even though FTPS runs at zero temperature, assuming that this does not change the spectral function significantly \cite{bauernfeind_fork_2017}.

The predicted spectral functions, both from our NN and MaxEnt, together with the original FTPS spectral function are shown in \pref{fig:res_ftps_spectral} (a).
The corresponding spectral function exhibits features typical for a correlated metal, a pronounced quasi-particle peak at $\omega=0$, together with a lower and upper Hubbard band which are both formed by localized atomic-like excitations.
Specificially, the upper Hubbard band consists of three distinct peaks, which are related to different atomic multiplets, see Ref.~\onlinecite{bauernfeind_fork_2017}. 

Both MaxEnt and the NN give a very good reconstruction of the quasiparticle peak.
The lower Hubbard peak (at $\omega \approx -2$\,eV) is also described reasonably well, even though MaxEnt broadens the peak considerably, and the NN slightly shifts the center of the peak to lower frequencies.
Neither MaxEnt nor the NN are able to capture the asymmetric, triangular shape of this peak, similar to what was observed for the example shown in Fig.~\ref{fig:res_special_spectral}(e).
The upper Hubbard band represents a significant challenge for both methods, due its three-peak structure with two rather sharp peaks at relatively high frequencies. MaxEnt essentially merges the two peaks at around 2.5 and 4\,eV into one broad peak and completely ignores the smaller peak at \SI{6}{eV}. Therefore, it significantly underestimates the width of the upper Hubbard band. The NN also fails to describe the true three-peak structure, but is able to produce more refined features, with the lowest of the three peaks partially resolved and a broader feature that stretches approximately across the energy range of the two higher energy peaks, such that the overall width of the upper Hubbard band is captured relatively well.
The uncertainty estimation of the NN correctly identifies the two Hubbard bands as problematic regions. 

We have also verified that the spectral functions obtained from a DFT+DMFT calculation for \SVO using the CT-QMC solver in combination with either MaxEnt or our NN are very similar to those shown in \pref{fig:res_ftps_spectral}, indicating that the differences are indeed related to the analytic continuation and not to the different solvers. This is consistent with the analysis presented in Ref.~\onlinecite{bauernfeind_fork_2017}.

In conclusion, our NN resolves the features of the FTPS spectral function better than MaxEnt. In particular, the uncertainty estimation proves to be very useful in identifying region where the prediction seems rather reliable and others where it might not be too accurate. 
However, the sharp features in the upper Hubbard band are not resolved by the NN. This is not too surprising, since the data set used for training and testing was designed to not contain any sharp peaks in the frequency region $|\omega| > \Omega_2 = 2.4$\,eV.
This constraint was mainly motivated by ``typical'' spectral functions for a variety of materials that can be found throughout the literature. However, most of these spectral functions have been obtained using MaxEnt and this apparently has biased our expectation on how a spectral function typically looks like. On the other hand, the application of the very recent FTPS solver was able to uncover further details in the spectral function of \SVO. This raises the question of whether the NN simply was not able to resolve these features because it had not seen comparable spectral functions previously, or whether it is related to the general difficulty to invert \pref{eq:forward_transform} for high frequencies, where the integral kernel becomes very small. This question is addressed in the next sub-section.

\subsection{Data recipe 2} \label{sec:res_recipe2}

Based on the spectral function obtained for \SVO using the FTPS solver, we now use the alternative data recipe 2 (see \pref{sec:data_recipes} and \pref{tab:data_recipes}) to generate a different set of spectral functions, which generally contain fewer, and thus more isolated, peaks that can still be rather narrow also at higher frequencies. 

\begin{figure}
    \centering
    \includegraphics[width=1\linewidth]{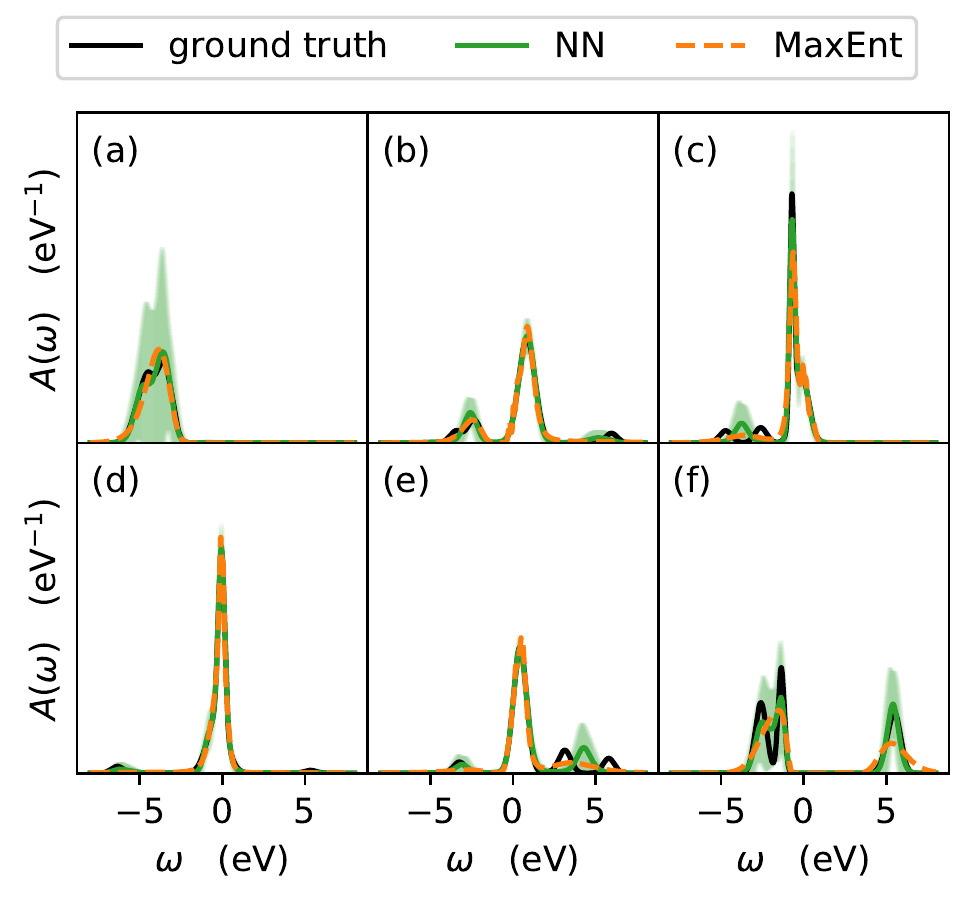}
    \caption{Examples of spectral functions constructed from data recipe 2. Each panel contains the ground truth, the NN prediction with the estimated uncertainty range, and the MaxEnt prediction. Predicted spectral functions are obtained with a noise level of $\sigma = 10^{-3}$ on $G_l$.}
    \label{fig:res_examples_spectral_nn2}
\end{figure}

\pref{fig:res_examples_spectral_nn2} shows a random selection of spectral functions generated with data recipe 2. The first observation is that these spectral functions indeed have more clearly separated peaks and less broad multi-peak structures. Nevertheless, the spectral functions still exhibit non-trivial features, such as shoulders or split peaks, where multiple Gaussians overlap.
The spectral function in panel (a) would represent a completely filled ``band insulator'', panels (b)-(e) would correspond to metallic systems, and panel (f) represents an insulator with partial filling. 

Both the NN and MaxEnt correctly capture the quasiparticle-like peaks around $\omega=0$ in panels (b)-(e) very well, but have difficulties to correctly describe high- or low-frequency peaks with small amplitude. Neither MaxEnt nor the NN can resolve the small two peak-structures at low and high frequencies in panels (c) and (e), respectively. In both cases, MaxEnt produces a very broad feature, whereas the NN creates a relatively sharp peak in between the two original peaks.
Case (a) seems rather well described, but the uncertainty estimation of the NN appears very high.
In panel (f), the two peak structure at negative frequencies is partially resolved by the NN and the high frequency peak is also described rather well, whereas MaxEnt just produces broad features.
Overall, the NN trained on data recipe 2 seems to perform well on the type of data it was trained for, generally leading to spectral functions that agree equally well or better with the ground truth than those obtained by MaxEnt.

Next, we apply the neural network trained on data recipe 2 (NN2) to the Green's function of \SVO obtained using the FTPS solver. The result is shown in \pref{fig:res_ftps_spectral} (b). Indeed the new NN2 results in sharper and more isolated features in the energy region of the upper Hubbard band, but is also not able to correctly resolve the three-peak structure. 
We note that this reflects the intrinsic problem of resolving the high-energy peaks from the kernel in \pref{eq:forward_transform} that all methods struggle with.
NN2 also gives higher uncertainty estimates and in fact performs quantitatively worse on this spectral function than NN1 (the original NN trained on data recipe 1), with a MAE of \num{.46} for NN2 compared to an MAE of \num{.33} for NN1.

\begin{figure}
    \centering
    \includegraphics[width=1\linewidth]{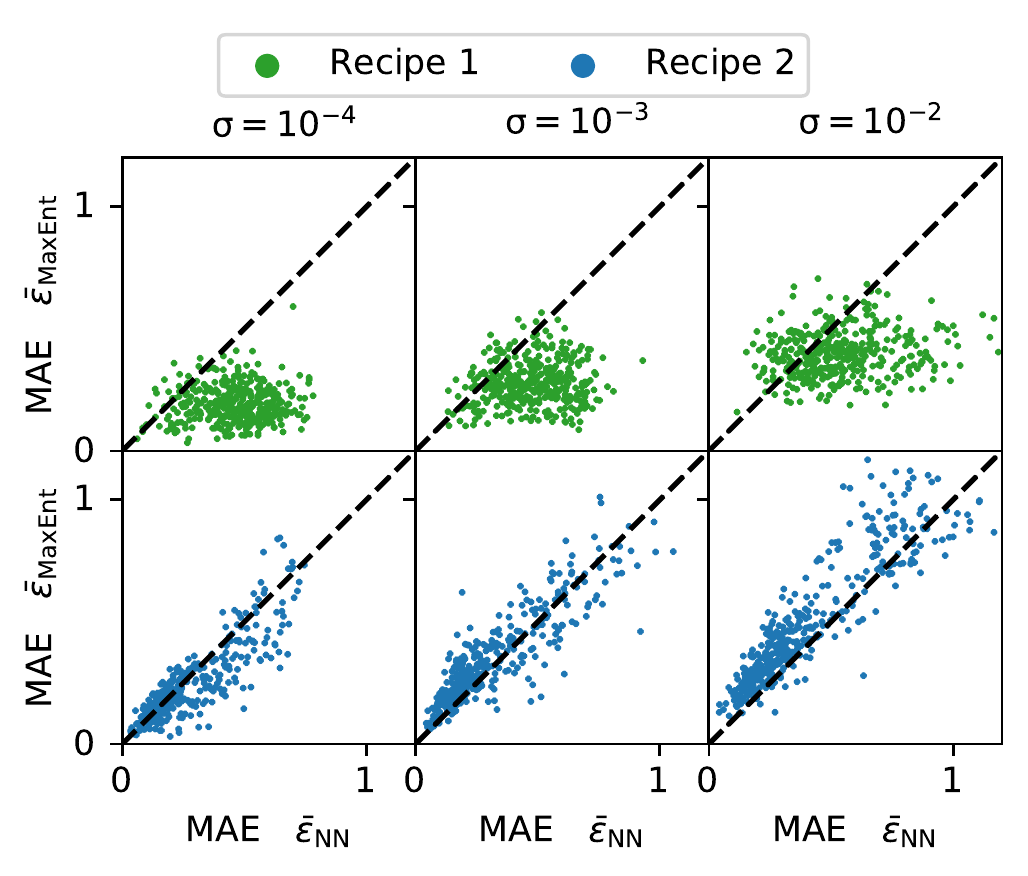}
    \caption{Cross-validation of the NNs trained on the two different data recipes. The top row shows the performance of NN2 on spectral functions generated according to data recipe 1. The bottom row shows the performance of NN1 on spectral functions generated according to data recipe 2. In both cases the MAE of the NN is compared to that achieved by MaxEnt on the same data.
    }
    \label{fig:res_crossvalid}
\end{figure}

The construction of a second data set also allows us to cross-validate the NNs by training it on one data recipe and testing it on the other type of data. The result is shown in \pref{fig:res_crossvalid}, where the performance of the two NNs on the respectively other data set is compared against MaxEnt for different noise levels. 
The top row indicates the performance of NN2 (the NN trained on data recipe 2) for predicting spectral functions from data recipe 1. The resulting average 
MAE is larger than that of MaxEnt for all noise levels (but becomes more comparable to that of MaxEnt for $\sigma=10^{-2}$. 
In contrast, NN1 (the NN trained on date recipe 1) performs much better on the unknown data set, the overall MAE is only slightly larger than that of MaxEnt for low noise levels, while for medium and high noise levels the performance is superior to that of MaxEnt.

This indicates that date recipe 1 is indeed more versatile than recipe 2 and the corresponding NN generally performs well, even on spectral functions different from the ones it was trained on. This also agrees with the performance on the previously unseen real-frequency spectral function of \SVO, and makes NN1 clearly preferable over NN2. 
\pref{app:uncertainty} contains an additional  quantitative cross-validation of the performance of the uncertainty estimation for both NNs. 

\section{Summary and conclusions}

In summary, we have implemented a machine-learning algorithm to address the problem of analytic continuation of imaginary time single-particle Green's functions obtained from DMFT using, e.g.,  QMC solvers.
Thereby, an important problem is the lack of real physical data, since there is no cheap and sufficiently accurate method to obtain the real-frequency spectral functions without resorting to analytic continuation, and existing methods for analytic continuation all suffer from the inversion of \pref{eq:forward_transform} being ill-conditioned.
Thus, similar to most previous work, we obtain data to train our multi-level residual NN by generating artificial spectral functions through a superposition of Gaussian peaks and then calculating the corresponding imaginary-time Green's function from the numerically stable forward-transform, \pref{eq:forward_transform}. 
However, the assumptions made on the possible shape of a typical spectral function and the specific form that is used for representation can introduce a bias in the predicted spectral functions. 
In this work, we have therefore put special emphasis on cross-validating the trained neural network on alternative data that does not necessarily match the training data. Furthermore, we have implemented an uncertainty estimation based on MC dropout, which helps to identify regions where the predicted spectral function might not be too accurate.

Overall, the multi-level residual NN used in this work performs very well, similar to previous work using machine learning for analytic continuation~\cite{arsenault_projected_2017, yoon_analytic_2018, fournier_artificial_2020, song_analytic_2020, xie_analytic_2022, huang_learned_2022}, and comparable or even better than the well established MaxEnt approach~\cite{skilling_maximum_1989, jarrell_bayesian_1996}. 
This also holds for spectral functions that are specifically designed to be different from the training data as well as for the spectral function obtained for \SVO without analytic continuation. (Except perhaps for the Lorentzian peaks, where MaxEnt outperforms the NN, even though the NN also still gives a reasonable prediction.)  
Our analysis for different noise levels during training and testing shows that the NN gives reliable predictions if the noise level of the test data is similar or lower than that used during training, but that an underestimation of the noise level during training significantly degrades the reliability of prediction. 

In particular, our benchmark using the real-frequency DMFT data for \SVO gives some important insights. First, it reveals that this spectral function indeed exhibits features that are not well represented in the training data. However, developing an improved data recipe is not straightforward. Our data recipe 2, which was specifically designed to incorporate more isolated and sharper peaks at higher excitation energies, did not result in a significant improvement for \SVO, and indeed turned out to be clearly less general than our original data recipe.
Finally, the corresponding analysis also indicates the fundamental limitations in correctly describing such higher energy features that result from the ill-conditioned kernel in \pref{eq:forward_transform}, which cannot easily be bypassed using machine learning.

In conclusion, machine learning shows great promise as an alternative to MaxEnt for analytic continuation. Based on the analysis presented here, it seems that currently the main limitation for further improvement is the lack of real training data and the biases that are potentially introduced by the way the artificial training data is generated. On the other hand, while a further optimization of the network architecture or training procedure might still be possible, it is unclear whether this would indeed also lead to a significant gain in performance for spectral functions that are substantially different from the training data.

\appendix*
\begin{acknowledgments}
We thank Ilan Bouquet for doing initial tests on applying machine learning to the problem of analytic continuation in his MSc project.
This research was supported by ETH Zurich and the NCCR MARVEL, a National Centre of Competence in Research, funded by the Swiss National Science Foundation (grant number 182892). Calculations have been performed on the cluster ``Piz Daint'' hosted by the Swiss National Supercomputing Centre and supported under project IDs s889 (User Lab) and
mr26 (MARVEL).
The Flatiron Institute is a division of the Simons Foundation.
\end{acknowledgments}

\appendix

\section{DFT+DMFT calculation details} \label{app:dft_dmft}

The DFT+DMFT calculations for the cubic perovskite \SVO presented in this work are performed without enforcing charge self-consistency between the DFT and DMFT part of the calculation and are based on nonmagnetic DFT calculations.
For the DFT+DMFT calculations presented in \pref{sec:effect_noise}, which are used to obtain a typical covariance matrix, we use the ``Vienna Ab-initio Simulation Package'' (VASP) \cite{kresse_ab_1993, kresse_efficient_1996} with the PBE exchange-correlation functional \cite{perdew_generalized_1996}.
The unit cell contains one formula unit of \SVO with a lattice constant of \SI{3.84}{\angstrom}.
A $9 \times 9 \times 9$ $k$-point mesh is used for Brillouin zone integration with tetrahedron smearing and a kinetic energy cutoff of \SI{450}{eV} is used for the plane wave basis set.
We then define the correlated subspace for the DMFT calculation using the local PAW projections from VASP onto the V-$t_{2g}$ orbitals and the ``PLO converter'' from the DFTTools library~\cite{schuler_charge_2018, aichhorn_triqsdfttools_2016}.
To obtain the spectral function for \SVO from DFT+DMFT discussed in \pref{sec:FTPS}, we use the \textsc{Quantum~ESPRESSO} package~\cite{Giannozzi_et_al:2009} with scalar-relativistic ultrasoft pseudopotentials~\cite{garrity_2014}, again using the PBE exchange-correlation functional. 
The lattice constant is \SI{3.86}{\angstrom}, where we again simulate a unit cell of one formula unit.
The energy cutoffs for the wave functions and charge density are set to \SI{60}{Ry} and \SI{720}{Ry}, respectively.
We use a $17 \times 17 \times 17$ $k$-point grid to sample the Brillouin zone, and a smearing of \SI{0.015}{Ry} in the Methfessel-Paxton scheme.
The correlated subspace is constructed for the same subspace as above, but using the \textsc{Wannier90} code \cite{Mostofi_et_al:2008} to construct maximally localized Wannier functions (MLWFs)~\cite{Marzari_et_al:2012}.

Based on these DFT calculations, we then run DMFT, where the local electron-electron interaction is described using the Kanamori Hamiltonian \cite{kanamori_electron_1963} with $\mathcal U = \SI{5.00}{eV}$ and $\mathcal J = \SI{.65}{eV}$.
For the DMFT QMC calculations in \pref{sec:effect_noise}, we use the continuous-time CT-HYB solver~\cite{seth_triqscthyb_2016} and an inverse temperature $\beta = 1/(k_\mathrm{B} T) = \SI{40}{eV^{-1}}$ to solve the effective impurity problem.
We sample the local Green's function with 30 Legendre coefficients, and first converge a solution with \num{8e7} solver cycles over 20 iteration, then run 200 independent runs from the same starting point with \num{2e7} cycles.
To solve the impurity problem using the fork tensor-product states (FTPS) solver \cite{bauernfeind_fork_2017} for \pref{sec:FTPS}, we first use Wannier interpolation to re-evaluate the Hamiltonian on a denser grid with $50$ $k$-points per dimension.
The calculation of the $T=0$ real-frequency spectral function was converged down to a broadening $\eta = \SI{0.05}{eV}$.
The number of bath sites was chosen automatically for the given broadening such that finite discretization effects from the bath fitting are avoided, yielding $N_b \approx 200$ per orbital.
The bond dimensions are set to 150 (300) for the impurity (bath-bath/bath-impurity) links both for the ground state search and for the time evolution.
We use a truncated weight of $t_w = \num{e-12}$ and $t_w = \num{e-9}$ for the ground state search and the time evolution, respectively.
The time evolution is performed up to $\SI{56}{eV^{-1}}$.

Both solvers are part of the TRIQS library \cite{parcollet_triqs_2015}. All the DMFT calculations are done with the publicly available solid\_dmft package \cite{solidmft}.

\section{(Cross-) validation of uncertainty estimation} \label{app:uncertainty}

\begin{table}
    \centering
    \caption{(Cross-) Validation of the uncertainty estimation of the two NNs on the two different data sets (recipes 1 and 2). The fraction of data-points for the pointwise absolute error that are within the uncertainty estimation, $\varepsilon(\omega) \leq 10 \sigma_\mathrm{dropout} (\omega)$, are specified for the different cases and different noise levels.}
    \begin{ruledtabular}
    \begin{tabular}{l|ccc}
    & $\sigma=10^{-4}$ & $\sigma=10^{-3}$ & $\sigma=10^{-2}$ \\ \hline
    NN1 on recipe 1 & 0.99 & 0.92 & 0.70 \\
    NN1 on recipe 2 & 0.94 & 0.88 & 0.72 \\
    NN2 on recipe 1 & 0.85 & 0.85 & 0.79 \\
    NN2 on recipe 2 & 0.98 & 0.96 & 0.90 \\
    \end{tabular}
    \end{ruledtabular}
    \label{tab:crossval_uncertainty}
\end{table}

To further validate the empirical approximate upper bound for the pointwise absolute error in terms of the standard deviation from the MC dropout, $\varepsilon(\omega) \leq 10 \sigma_\mathrm{dropout} (\omega)$, we evaluate the fraction of data points $\varepsilon(\omega)$ that respect this bound and compare this fraction for the differently trained NNs and different noise levels. The result obtained for 1000 test spectral functions in each case is summarized in \pref{tab:crossval_uncertainty}.

Generally, it seems that the accuracy of this upper bound decreases with increasing noise level, indicated by the smaller fraction of points that respect this limit. This decrease is more pronounced for NN1 but also occurs for NN2. Furthermore, for a low noise level of $\sigma=10^{-4}$, the lower fraction of points within the upper bound observed for NN2 applied on spectral functions generated according to data recipe 1 indicates a certain degree of overfitting, as already observed in the cross-validation of the MAE (Fig.~\ref{fig:res_crossvalid}).

\bibliography{bibfile}

\end{document}